# Unmasking Biases and Navigating Pitfalls in the Ophthalmic Artificial Intelligence Lifecycle: A Review

## Authors

Luis Filipe Nakayama, MD, [1,2] João Matos, MSc, [1,3,4] Justin Quion, MD, [1] Frederico Novaes, MD, [2] William Greig Mitchell, MBBS MPH, [5] Rogers Mwavu, [6] Ju-Yi Ji Hung, MD, PhD, [7, 8] Alvina Pauline dy Santiago, MD, [9, 10, 11, 12] Warachaya Phanphruk, MD, [13] Jaime S. Cardoso, PhD, [3,4] Leo Anthony Celi, MD, MS, MPH, [1,14,15]

## Affiliations

1. Laboratory for Computational Physiology, Institute for Medical Engineering and Science, Massachusetts Institute of Technology, Cambridge, MA, USA
2. Department of Ophthalmology, São Paulo Federal University, São Paulo, SP, Brazil
3. Faculty of Engineering, University of Porto (FEUP), Porto, Portugal
4. Institute for Systems and Computer Engineering, Technology and Science (INESCTEC), Porto, Portugal
5. Department of Ophthalmology, Royal Victorian Eye and Ear Hospital, Melbourne, Australia
6. Department of Information Technology, Mbarara University of Science and Technology, Mbarara, Uganda

7. Department of Ophthalmology, Byers Eye Institute at Stanford, California, USA

8. Department of Computer Science and Information Engineering, National Taiwan University, Taiwan

9. Pediatric Ophthalmology, University of the Philippines Manila College of Medicine, Manila, Philippines

10. Pediatric Ophthalmology Department of Ophthalmology and Visual Sciences, Philippine General Hospital, Manila, Philippines

11. Pediatric Ophthalmology Eye and Vision Institute, The Medical City, Manila, Philippines

12. Pediatric Ophthalmology International Eye Institute St Luke's Medical Center, Quezon City, Philippines

13. Department of Ophthalmology, Faculty of Medicine, Khon Kaen University, Khon Kaen, Thailand

14. Harvard TH Chan School of Public Health, Department of Biostatistics, Boston, MA, USA

15. Beth Israel Deaconess Medical Center, Department of Medicine, Boston, MA, USA

## Corresponding author

Luis Filipe Nakayama ([luisnaka@mit.edu](luisnaka@mit.edu))

## Author contributions

All authors contributed to writing the manuscript and approved the final version. JC and LAC reviewed the paper and supervised the work.

## Data availability statement

No new data were generated or analyzed in support of this research.

## Funding Statement

The authors received no specific funding for this work.

## Competing interests

LAC is the Editor-In-Chief of PLOS Digital Health.

## Abstract

Over the past two decades, exponential growth in data availability, computational power, and newly available modeling techniques has led to an expansion in interest, investment, and research in Artificial Intelligence (AI) applications. Ophthalmology is one of many fields that seek to benefit from AI given the advent of telemedicine screening programs and the use of ancillary imaging. However, before AI can be widely deployed, further work must be done to avoid the pitfalls within the AI lifecycle. This review article breaks down the AI lifecycle into seven steps: data collection; defining the model task; data pre-processing and labeling; model development; model evaluation and validation; deployment; and finally, post-deployment evaluation, monitoring, and system recalibration and delves into the risks for harm at each step and strategies for mitigating them.

## Author summary

In recent years, the surge in data availability, computational power, and AI techniques has sparked interest in using AI in fields like ophthalmology. However, before widespread AI deployment can happen, we must carefully navigate its lifecycle, comprising seven key steps: data collection, task definition, data preparation, model development, evaluation, deployment, and post-deployment monitoring.

This review article stands out by identifying potential pitfalls at each stage and offering actionable strategies to address them. Our article serves as a guide for harnessing AI effectively and safely in ophthalmology and related fields.

## Introduction

The use of computers simulating humans in performing cognitive tasks was first described in 1943[1], but the term "Artificial Intelligence" (AI) would not be coined until 1956. At that time, the fundamental premise was that "every aspect of learning or any other feature of intelligence can, in principle, be so precisely described that a machine can be made to simulate it"[2]. Over the last two decades, the exponential growth of data availability and computational power has led to an expansion in interest, investment, and research in AI applications[3].

Machine Learning (ML) is a subfield of AI that focuses on "teaching" computers how to make predictions, classify or optimize some function from datasets. While traditional ML requires a fair amount of work with feature engineering[4], Deep Learning (DL) is a ML technique where there are no predefined associations between variables, allowing for computers to utilize multiple processing layers and multiple levels of abstraction to produce predictions, classifications or optimal policies[5,6]. DL models have been used in several healthcare processes, including electronic medical records analysis, predicting patient outcomes, and image analysis[7].

In ophthalmology, the advent of telemedicine screening programs and the widespread use of ancillary imaging examinations have created large volumes of ophthalmic data, enabling the development of AI algorithms[8]. The ophthalmic community has since witnessed the application of AI to a wide range of diagnostic examinations with specialist-level performance[9–12]. Previous studies report the ability of AI-based algorithms to detect diabetic retinopathy[13–15], diabetic macular edema[16], retinopathy of prematurity[17,18], age-related macular degeneration[19–22], glaucoma[23–26], and uveitis[27], among others.

However, deployed models in clinical practice have failed to meet expectations, as seen in COVID-19 prediction, sepsis detection, and diabetic retinopathy screening algorithms[28–30]. One huge concern is the unintentional encryption and propagation of biases at every stage of the AI lifecycle from data collection to post-deployment and recalibration[31]. In addition to data and algorithmic bias, other issues in the life cycle include reliance on irrelevant evaluation metrics, and a lack of post-deployment studies[28,29,32]. Identifying pitfalls at every step of the AI pipeline is fundamental to breaking the cycle of health inequities.

## The AI Lifecycle in Ophthalmology

In this article, we divided the ophthalmic AI lifecycle into seven steps consisting of data collection; defining the model task; data pre-processing and labeling; model development; model evaluation and validation; deployment; and finally, post-deployment evaluation, monitoring, and system recalibration. In every step, we identify pitfalls and challenges. (Figure 1)

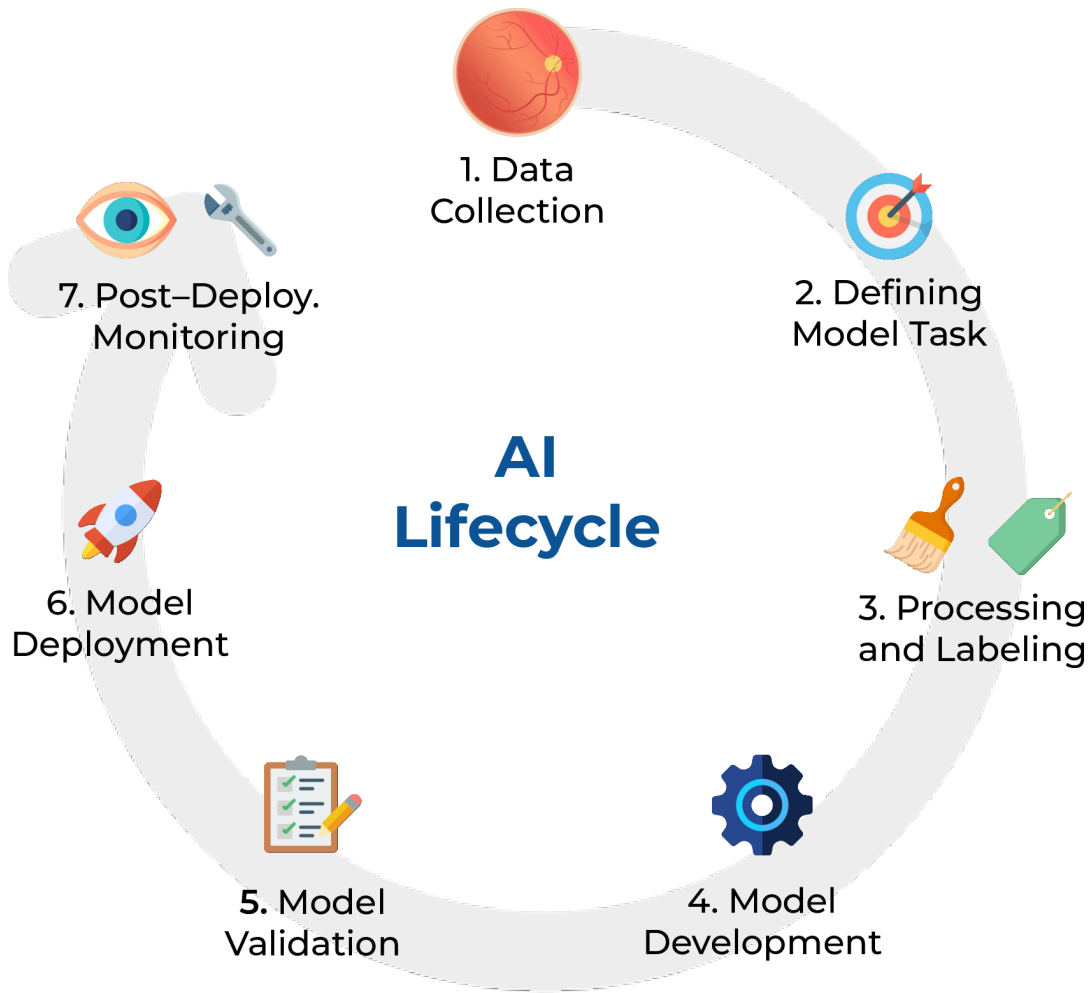

**Figure 1.** The seven steps of the Ophthalmologic AI lifecycle (Diagram was designed using icons from *Flaticon.com*).

Data collection

Clinical and imaging data needs to be collected to train AI algorithms. During this process, it is important to understand and document the population characteristics, data modalities, and included features and variables. In ophthalmology, images are collected during clinical practice and in screening programs. (Table 1)

Assembling new datasets is a laborious and expensive process. As a result, the ML community has mostly relied on publicly available datasets[33,34]. When employing these datasets, it is necessary to consider the data generation process, including but not limited to the demographic diversity of the patients, the examination criteria, and the disease distribution within the cohort. It is vital that the target population is adequately represented in the data during the development of ophthalmological models that detect ocular abnormalities in real-world settings[9].

**Table 1:** Ophthalmological data modalities, subtypes, and diagnostic aims.

| Data modality | | Aim |
|---|---|---|
| Imaging | **Retinal fundus photo** | *Diabetic retinopathy* |
| | | *Age-Related Macular Degeneration* |
| | | *Retinopathy of Prematurity* |
| | | *Glaucoma* |
| | **Optical Coherence Tomography** | *Diabetic Macular Edema* |
| | | *Age-Related Macular Degeneration* |
| | | *Macular diseases* |
| | **Anterior Segment photos** | *Corneal keratitis* |

|  |  | Cataract |
|  | **Topography** | Keratoconus |
|  | **Visual Field** | Glaucoma |
|  |  | Optic Neuritis |
| Tabular | **Electronic Medical Records** | Myopia |
|  |  | Visual acuity |
|  | **Clinical notes** | Ocular lesions characteristics |

## Pitfalls

**Unbalanced data sources**

Representation is an important attribute to consider when assessing data to train algorithms. The use of overly restrictive inclusion criteria can result in the exclusion of several subgroups, while applying too broad criteria can lead to an increased number of "false positive" patients within a cohort[35]. Moreover, underrepresented populations traditionally are not included in datasets, and even in perfectly representative data, historical biases can lead to misrepresentation[36].

To date, ophthalmic datasets used in AI research are significantly imbalanced with no retinal imaging datasets from middle-low or low-income countries[34]. This imbalance has significant implications for the development of generalizable models.

In ophthalmology, including only images from screening programs and patients from a pre-specified population, demographic, or health system will introduce bias and lead to a disparity in ML performance across subpopulations[37]. While it may be impossible to fully represent all subpopulations, efforts to understand the social patterning of the data generation process and engage with marginalized populations to develop strategies for inclusive data collection and analysis are required. At the minimum, awareness of the limitations of datasets should be reported.

**Image acquisition, protocols, and image quality**

The image acquisition process involves the implementation of an imaging protocol, which can vary depending on the specific disease being investigated. In the case of diabetic retinopathy screening, it initially entailed capturing seven 30-degree fundus photos per eye, but evolved to either two 45-degree photos or a single image, depending on the screening program[38].

Image quality standards for datasets were established by experts from a handful of high-income countries, with little, if any, contribution from low- and middle-income countries. Such data standards may discourage creators of small or "low quality" datasets from contributing to data repositories. Additionally, the same rich countries have the infrastructure and resources to create, curate, and maintain data warehouses and AI pipelines. Setting high standards for image quality in ophthalmological examinations may lead to better model performance but trades off real-world applicability[30]. It is important to ensure that initiatives to establish dataset standards involve more inclusive teams that consider the realities in most countries.

**Spectrum bias**

Spectrum bias occurs when the diseases studied within the dataset are not fully representative of the spectrum of disease severity in the target population, leading to poor model performance[39]. Ophthalmological datasets skewed toward mild cases of diabetic retinopathy may lead to algorithms that are better at detecting mild cases while missing severe cases and vice versa [34,39].

Therefore, it's crucial to consider the model's performance across the full spectrum of the disease. The AI model should be trained and validated on diverse patient data, including different ages, ethnicities, disease stages, and comorbidities, to ensure applicability and reliability in ophthalmology.

## Defining the model task

In ophthalmology, understanding the target populations' clinical context, examination modalities, and disease prevalence and distribution is essential. Moreover, a comprehensive understanding of the socioeconomic context, the healthcare delivery system, and its resources are important for identifying ocular conditions to target and suitable tasks for training algorithms[40].

## Pitfalls

**Defining the appropriate disease target**

The prevalence of ocular diseases varies significantly across geography and is shaped by several demographic and socioeconomic factors. Defining the disease target must be tailored according to the specific population and the healthcare context. While AI has seen applications in various ophthalmological diseases, diabetic retinopathy screening has been the most extensively explored to date[41]. While diabetic retinopathy is a global epidemic[2], there are many other causes of preventable blindness such as unaddressed refractive errors, affecting 914 million people worldwide; cataracts, affecting 94 million individuals; Age-related macular degeneration, affecting 8 million; and glaucoma, affecting 7.7 million people [42]. Several other critical causes of blindness, such as keratoconus, infectious keratitis, and deficiency neuropathies, remain uncharted territories for AI technologies. These conditions are especially prevalent in economically disadvantaged regions where access to diagnostic laboratory services is limited[6]. Resources must be allocated based on the specific burden of ocular diseases in different regions.

**Data sparsity bias**

Data sparsity bias can manifest when a dataset lacks sufficient case numbers for specific combinations of exposure and outcomes[43]. This limitation often occurs despite the use of large databases, resulting in a restricted sample of cases, risk factors, variables, and outcomes.

Traditionally, two approaches are employed to mitigate this: adjustment and penalization. While these methods may be effective, a deep understanding of the clinical context and underlying social determinants is necessary[43]. Failing to incorporate this information in the models will exacerbate existing biases. Researchers should actively seek to enhance data collection efforts, particularly in underrepresented or vulnerable populations.

## Data preprocessing and labeling

Data preprocessing is an important step in building AI algorithms. Computer vision models, especially with traditional image analysis techniques, require preprocessing, consisting of image enhancement, filtering, and normalization. Additionally, data augmentation can enhance the size and quality of training data sets. This improves the robustness of models, particularly when dealing with limited data.

In ophthalmology, the preprocessing stage is even more critical, given the number of distinct capturing processes and retinal fundus photography devices. Ensuring accurate and reliable results necessitates accounting for the inherent technical differences among images.

## Pitfalls

**Missingness handling bias**

Missing data are due to many reasons, including human and machine limitations, errors during data collection, limited accessibility to clinical screening and examinations, and respondents' refusal to participate in research[44]. In ophthalmology datasets, missing information on comorbidities, demographics, and/or ophthalmological examination data is common [34]. Several methods exist for dealing with missingness, such as deleting instances with missing values or imputing the missing data using estimated values. However, the choice of imputation method should be made carefully, as different methods may introduce additional layers of bias.

To enhance the quality of datasets during development and reduce potential biases, it is essential to engage a diverse team of ophthalmologists, statisticians, data scientists, and even patient representatives as multi-stakeholder discussions will lead to more informed decisions [44,45].

**Labeling**

In supervised machine learning, accurate and reliable labels are necessary for training algorithms to predict diagnoses and outcomes. The labeling process in ophthalmology is complex and challenging, involving varying application of grading criteria and standards. Errors and inconsistencies in the labeling process can lead to biases in the models, ultimately impacting the performance and reliability of algorithms[46].

To address this issue, alternatives such as graders' consensus and expert adjudication can be employed to enhance the reliability of ophthalmological labeling[46]. The use of weakly supervised learning techniques offers a promising approach to improve the annotation process, making labeling more efficient and effective[47,48].

## Model development

The model development phase consists of data curation and representation, and knowledge creation and validation. Avoiding biases during modeling is an arduous task[49,50] and requires an in-depth understanding of data disparities, clinical confounders, and the socioeconomic context[50].

## Pitfalls

**Flawed feature engineering or selection**

The relevance of features when predicting or optimizing outcomes may vary across patient subgroups. When dealing with tabular data or traditional imaging techniques that require feature engineering and selection, it is essential to acknowledge the risk of both under- and over-reliance on certain features. Not all features that can be extracted from EHRs or imaging exams are relevant and their inclusion in modeling may worsen task performance. The input of domain experts in this complex process is paramount.

**Diagnostic suspicion bias**

The knowledge of a patient's prior exposure or pre-existing conditions can introduce diagnostic suspicion bias in datasets[51,52]. For instance, when a patient has multiple known comorbidities, it may lead to a more extensive work-up, affecting the data collected for model development. Careful consideration is necessary when incorporating variables related to the outcome, such as the frequency of optical coherence tomography (OCT) and retinal fundus photos, as it could inadvertently bias the model towards specific diseases and risk factors.

**Data leakage**

Data leakage occurs when information from the training data is also captured in the validation cohort, or information about the outcome is reflected in the features, resulting in a falsely accurate model performance[53]. Notable examples from previously published models include using an antibiotic prescription to predict a sepsis diagnosis, or the blessing of a hospital chaplain to predict mortality. However, data leakage may be subtle and more difficult to detect[54].

In the context of ophthalmology, data leakage can occur when 3D volumetric OCT images are split on a per-image basis[53]. In this scenario, valuable spatial information from the volumetric data may inadvertently influence the model's predictions in ways that compromise the model's true ability to generalize to new data.

**Shortcuts**

Model shortcuts pertain to features that are learned during model development that are not clinically related to a prediction or classification, such as the hospital where a patient is seen or the medical equipment used for imaging[55]. Studies have shown that algorithms can perform seemingly impossible tasks, such as determining a patient's sex from a retinal fundus photography[56] or identifying race from a chest X-ray[57]. Such findings raise concerns about the use of such "invisible" features for diagnosis and treatment recommendation, rather than those pertaining to the clinical features of the disease.

In ophthalmology, small datasets, image details, and class imbalance can contribute to shortcut learning and measurement bias[31,58]. A thorough data analysis, careful interpretation of results, investigation of model explainability, and generalizability of a test model are needed to assess the impact of shortcut features [55].

## Model evaluation and validation

AI algorithms are usually tested on external datasets to evaluate their real-world performance. Typical metrics include accuracy, sensitivity, specificity, the receiver operating characteristic (ROC) curve, and F1-score. It is crucial to choose the appropriate metrics and to perform a thorough analysis of downstream consequences of errors.

Pitfalls

**Wrong evaluation metric**

Using inappropriate evaluation metrics can hide the underperformance of models in certain patient subgroups and widen outcome disparities[31]. To ensure algorithmic fairness, it is imperative to assess the performance across marginalized cohorts [59]. It is also essential to acknowledge that all evaluation metrics are only estimations of a construct and, therefore, may not fully capture the true extent of algorithmic fairness[55]. Metrics relying solely on accuracy, or other characteristics based on historical data, run the risk of perpetuating health inequities present in the data[59].

**Wrong evaluation method**

Evaluation bias arises when the methods used to assess model performance themselves are biased. Inadequate external and benchmark datasets that fail to represent the population accurately can result in a biased evaluation of the model's performance. To improve the robustness and generalizability of algorithms, and avoid shortcut learning, it is important to use external validation datasets that are representative of the target population [31,55].

## Model deployment

The deployment of AI systems in real-world settings is the culmination of understanding the data characteristics and the social patterning of the data generation process, model task definition, data preprocessing, and model evaluation. Deployment and post-deployment monitoring and recalibration may introduce additional bias and require as much risk mitigation.

### Pitfalls

**Defining the model threshold**

Defining the decision thresholds depends on the model's purpose (e.g., screening, diagnosis, triage) and the health system it is to be deployed in.

For example, in diabetic retinopathy screening, setting lower thresholds may increase the number of false positive results and unnecessary referral cases and workup. While this could lead to a higher sensitivity for detecting potential cases, it will also burden the healthcare system. Conversely, setting higher thresholds might reduce false positives but could lead to missed cases, compromising the sensitivity of the screening program and potentially delaying necessary interventions.

Comprehending the setting in which the model will be deployed is pivotal to its impact[39]. Deploying AI models, typically developed on data from well-resourced health systems, in limited-resource settings necessitates careful consideration of the available resources within the healthcare system [15]. To strike a balance between sensitivity and specificity, it is essential to consider factors such as the prevalence of the condition, the healthcare resources available, and the anticipated impact on patient outcomes. Multi-stakeholder engagement is crucial in this process.

**Covariate and Dataset shift**

Covariate shifts can arise when there are changes in the distribution of the input variables between the training and deployment phase.  In the field of ophthalmology, the generalization capabilities of AI models can be compromised due to shifts that occur when transitioning from the training dataset to the actual population they are being deployed to. These shifts can be attributed to various factors, including changes in patient demographics, changes in disease prevalence, differences in ophthalmological assessment timing, or variations in equipment use. Furthermore, population drift may occur as a result of a practice pattern change or enactment of new health policies[60,61].

## Post-Deployment Evaluation, Monitoring, and System Recalibration

It is required to conduct post-deployment monitoring of algorithms and analyze their impact on clinical outcomes with special attention to historically disenfranchised groups within the health system. Continuous measurement of the impact disaggregated across patient subgroups and recalibration of the model as necessary will ensure equitable benefit of AI systems[62,63]. Moreover, it is important to preempt unintended consequences, including heightening social stigma when outcomes among certain groups are highlighted, before integrating the algorithm into clinical practice.

## Pitfalls

**Model updates**

The decision of when and how to update AI systems can be a potential source of bias. The procedures for updating AI systems may differ across different settings, which can lead to differential performance and impact on patient outcomes. Consequently, it becomes essential to thoughtfully assess the potential biases that may arise during the updating process and implement appropriate measures to mitigate them. Timely updates to AI systems are essential for preserving their relevance as new clinical insights and practice patterns emerge.

## Conclusion

The implementation of AI systems in healthcare holds great promise for enhancing clinical decision-making. Ophthalmology, in particular, employs imaging ancillary examinations for clinical practice, and telemedicine screening programs enable the development of automated systems.

However, it is important to be aware that bias can be introduced at every step of the AI lifecycle. To address this concern we propose a seven-step process that encompasses data collection; defining the model task; data pre-processing and labeling; model development; model evaluation and validation; deployment; and finally, post-deployment evaluation, monitoring, and system recalibration. Understanding the impact of data disparities, modeling decisions, and evaluation metrics is vital to developing models that are accurate, equitable, and clinically relevant.

Collaboration between ophthalmologists, data scientists, social scientists, community representatives and other stakeholders is crucial to this endeavor. In resource-limited settings, the tailoring of AI systems to address practical challenges becomes essential for successful real-world deployment. Moreover, the continuous monitoring of AI systems post-deployment is paramount to identify any biases that may emerge over time and ensure ongoing fairness and transparency. Vigilance in identifying and addressing biases, combined with preemptive measures to prevent unintended consequences, enables AI systems to be leveraged responsibly and ethically in clinical practice.

100650